\documentclass[12pt]{article}
\usepackage{enumerate}
\usepackage{amsfonts}
\usepackage{amsmath}
\usepackage{amssymb}
\usepackage{amsthm}

\def\H{\mathcal{H}}

\def\S{\mathfrak{S}}

\def\T{\mathfrak{T}}
\def\B{\mathfrak{B}}

\newcommand{\supp}{\mathrm{supp}}

\newcommand{\id}{\mathrm{Id}}
\newcommand{\Tr}{\mathrm{Tr}}

\newcommand{\shs}{\hspace{1pt}}

\newcounter{defin}  \newcounter{lemma}  \newcounter{theorem}
\newcounter{property} \newcounter{corol}  \newcounter{remark} \newcounter{example}

\newenvironment{lemma}{\par\refstepcounter{lemma}
     \textbf{Lemma \thelemma.} }{\rm\par}

\newenvironment{property}{\par\refstepcounter{property}
     \textbf{Proposition \theproperty.}\ }{\rm\par}

\newenvironment{remark}{\par\refstepcounter{remark}
     \textbf{Remark \theremark.}}{\rm\par}

\begin{document}

\title{Continuity bounds for information characteristics of quantum channels depending on  input dimension}
\author{M.E. Shirokov\footnote{Steklov Mathematical Institute, RAS, Moscow, email:msh@mi.ras.ru}}
\date{}
\maketitle

\begin{abstract}
We show how to use properties of the quantum conditional mutual information to obtain continuity bounds for information characteristics of quantum channels depending on their input dimension.

First we prove tight estimates for variation of the output Holevo quantity with respect to simultaneous variations of a channel and of an input ensemble.

Then we obtain tight continuity bounds for output conditional mutual information for a single channel and for $n$ copies of a channel.

As a result tight and close-to-tight continuity bounds for basic capacities of quantum channels depending on the input dimension are obtained. They  complement the Leung-Smith continuity bounds depending on the output dimension.
\end{abstract}

\section{Introduction and preliminaries}

Leung and Smith obtained in \cite{L&S} continuity bounds for basic capacities of  quantum channels depending of their output dimension.
The appearence of the output dimension in these and some other continuity bounds for information characteristics of a quantum channel is natural, since such characteristics are typically expressed via entropic quantities of output states of a channel (so,  application of the Fannes type continuity bounds to these quantities gives the output dimension factor \cite{ A&F,Aud,Fannes, W-CB}).

In this paper we show how to use properties of the quantum conditional mutual information to obtain continuity bounds for information characteristics of quantum channels depending on their \emph{input dimension}.

We begin  with the output Holevo quantity $\chi_{\Phi}(\{p_i,\rho_i\})$ -- the Holevo quantity of ensemble $\{p_i,\Phi(\rho_i)\}$ obtained by action of a channel $\Phi$ on a given ensemble of input states. We obtain tight estimates for variation of $\chi_{\Phi}(\{p_i,\rho_i\})$ with respect to simultaneous variations  of a channel $\Phi$ and of an input ensemble $\{p_i,\rho_i\}$.\smallskip

Then  we obtain tight continuity bounds for the output conditional mutual information $I(B\!:\!D|C)_{\Phi\otimes\id_{CD}(\rho)}$ with respect to simultaneous variations  of a channel  $\Phi:A\rightarrow B$ and of an input state  $\rho_{ACD}$.  We will also derive tight continuity bound for the function $\Phi\mapsto I(B^n\!:\!D|C)_{\Phi^{\otimes n}\otimes\id_{CD}(\rho)}$ for any natural $n$ by using the Leung-Smith telescopic trick.\smallskip

The above results are applied to obtain tight and close-to-tight continuity bounds for basic capacities of quantum channels depending on their input dimension. They complement the above-mentioned Leung-Smith continuity bounds (depending on the output dimension).\medskip\medskip\medskip

Let $\mathcal{H}$ be a finite-dimensional or separable infinite-dimensional Hilbert space,
$\mathfrak{B}(\mathcal{H})$ the algebra of all bounded operators with the operator norm $\|\cdot\|$ and $\mathfrak{T}( \mathcal{H})$ the
Banach space of all trace-class
operators in $\mathcal{H}$  with the trace norm $\|\!\cdot\!\|_1$. Let
$\mathfrak{S}(\mathcal{H})$ be  the set of quantum states (positive operators
in $\mathfrak{T}(\mathcal{H})$ with unit trace) \cite{H-SCI,N&Ch,Wilde}.

Denote by $I_{\mathcal{H}}$ the identity operator in a Hilbert space
$\mathcal{H}$ and by $\id_{\mathcal{\H}}$ the identity
transformation of the Banach space $\mathfrak{T}(\mathcal{H})$.\smallskip

If quantum systems $A$ and $B$ are described by Hilbert spaces  $\H_A$ and $\H_B$ then the bipartite system $AB$ is described by the tensor product of these spaces, i.e. $\H_{AB}\doteq\H_A\otimes\H_B$. A state in $\S(\H_{AB})$ is denoted $\omega_{AB}$, its marginal states $\Tr_{\H_B}\omega_{AB}$ and $\Tr_{\H_A}\omega_{AB}$ are denoted respectively $\omega_{A}$ and $\omega_{B}$.\smallskip

A \emph{quantum channel} $\,\Phi$ from a system $A$ to a system
$B$ is a completely positive trace preserving linear map
$\mathfrak{T}(\mathcal{H}_A)\rightarrow\mathfrak{T}(\mathcal{H}_B)$,
where $\mathcal{H}_A$ and $\mathcal{H}_B$ are Hilbert spaces
associated with the systems $A$ and $B$ \cite{H-SCI,N&Ch,Wilde}.\smallskip

For any  quantum channel $\,\Phi:A\rightarrow B\,$ Stinespring's theorem implies the existence of a Hilbert space
$\mathcal{H}_E$ and of an isometry
$V:\mathcal{H}_A\rightarrow\mathcal{H}_B\otimes\mathcal{H}_E$ such
that
\begin{equation}\label{St-rep}
\Phi(\rho)=\mathrm{Tr}_{E}V\rho V^{*},\quad
\rho\in\mathfrak{T}(\mathcal{H}_A).
\end{equation}
The quantum  channel
\begin{equation}\label{c-channel}
\mathfrak{T}(\mathcal{H}_A)\ni
\rho\mapsto\widehat{\Phi}(\rho)=\mathrm{Tr}_{B}V\rho
V^{*}\in\mathfrak{T}(\mathcal{H}_E)
\end{equation}
is called \emph{complementary} to the channel $\Phi$
\cite[Ch.6]{H-SCI}.\smallskip

The set of quantum channels  is typically equipped with the metric induced by
the diamond norm
$$
\|\Phi\|_{\diamond}\doteq \sup_{\rho\in\T(\H_{AR}),\|\rho\|_1=1}\|\Phi\otimes \id_R(\rho)\|_1,\vspace{-5pt}
$$
on the set of all completely positive maps from $\T(\H_A)$ to $\T(\H_B)$. This norm  coincides with the norm of complete boundedness of the dual map $\Phi^*:\B(\H_B)\rightarrow\B(\H_A)$ to the map $\Phi$ \cite{H-SCI,L&S,Wilde}.\smallskip

For our purposes it is more convenient to use the equivalent metric on the set of channels called \emph{Bures distance} defined for given channels $\Phi:A\rightarrow B$  and $\Psi:A\rightarrow B$ as follows (cf.\cite{Kr&W})
$$
\beta(\Phi,\Psi)=\inf\|V_{\Phi}-V_{\Psi}\|,
$$
where the infimum is over all common Stinespring representations:
\begin{equation}\label{c-S-r}
\Phi(\rho)=\Tr_E V_{\Phi}\rho V^*_{\Phi}\quad\textrm{and}\quad\Psi(\rho)=\Tr_E V_{\Psi}\rho V^*_{\Psi}.
\end{equation}
It is proved in \cite{Kr&W} that
\begin{equation}\label{DB-rel}
\textstyle\frac{1}{2}\|\Phi-\Psi\|_{\diamond}\leq\beta(\Phi,\Psi)\leq\sqrt{\|\Phi-\Psi\|_{\diamond}},
\end{equation}
which shows the equivalence of the Bures distance and the diamond norm distance on the set of all channels between given quantum systems.\smallskip

The \emph{von Neumann entropy} $H(\rho)=\mathrm{Tr}\eta(\rho)$ of a
state $\rho\in\mathfrak{S}(\mathcal{H})$, where $\eta(x)=-x\log x$,
is a concave nonnegative lower semicontinuous function on the set $\mathfrak{S}(\mathcal{H})$, it is continuous if and only if $\,\dim\H<+\infty$ \cite{H-SCI,L-2,Wilde}.\smallskip

The \emph{quantum relative entropy} for two states $\rho$ and
$\sigma$ in $\mathfrak{S}(\mathcal{H})$ is defined as follows
$$
H(\rho\,\|\shs\sigma)=\sum\langle
i|\,\rho\log\rho-\rho\log\sigma\,|i\rangle,
$$
where $\{|i\rangle\}$ is the orthonormal basis of
eigenvectors of the state $\rho$ and it is assumed that
$H(\rho\,\|\sigma)=+\infty$ if $\,\mathrm{supp}\rho\shs$ is not
contained in $\shs\mathrm{supp}\shs\sigma$ \cite{H-SCI,L-2,Wilde}.\smallskip

The \emph{quantum mutual information} of a state $\,\omega_{AB}\,$ of a
bipartite quantum system  is defined as follows
\begin{equation}\label{mi-d}
I(A\!:\!B)_{\omega}=H(\omega_{AB}\shs\Vert\shs\omega_{A}\otimes
\omega_{B})=H(\omega_{A})+H(\omega_{B})-H(\omega_{AB}),
\end{equation}
where the second expression  is valid if $\,H(\omega_{AB})\,$ is finite \cite{L-mi,Wilde}.\smallskip

Basic properties of the relative entropy show that $\,\omega\mapsto
I(A\!:\!B)_{\omega}\,$ is a lower semicontinuous function on the set
$\S(\H_{AB})$ taking values in $[0,+\infty]$. It is well known that
\begin{equation}\label{MI-UB}
I(A\!:\!B)_{\omega}\leq 2\min\left\{H(\omega_A),H(\omega_B)\right\}
\end{equation}
for any state $\omega_{AB}$ and that
\begin{equation}\label{MI-UB+}
I(A\!:\!B)_{\omega}\leq \min\left\{H(\omega_A),H(\omega_B)\right\}
\end{equation}
for any separable state $\omega_{AB}$ \cite{MI-B,Wilde}.\smallskip

The \emph{quantum conditional mutual information} of a state $\omega_{ABC}$ of a
tripartite finite-dimensional system  is defined as follows
\begin{equation}\label{cmi-d}
    I(A\!:\!B|C)_{\omega}\doteq
    H(\omega_{AC})+H(\omega_{BC})-H(\omega_{ABC})-H(\omega_{C}).
\end{equation}
This quantity plays important role in quantum
information theory \cite{D&J,Wilde}, its nonnegativity is a basic result well known as \emph{strong subadditivity
of von Neumann entropy} \cite{Simon}. If system $C$ is trivial then (\ref{cmi-d}) coincides with (\ref{mi-d}).\smallskip

In infinite dimensions formula (\ref{cmi-d}) may contain the uncertainty
$"\infty-\infty"$. Nevertheless the
conditional mutual information can be defined for any state
$\omega_{ABC}$ by one of the equivalent expressions
\begin{equation}\label{cmi-e+}
\!I(A\!:\!B|C)_{\omega}=\sup_{P_A}\left[\shs I(A\!:\!BC)_{Q_A\omega
Q_A}-I(A\!:\!C)_{Q_A\omega Q_A}\shs\right],\; Q_A=P_A\otimes I_{BC},\!
\end{equation}
\begin{equation}\label{cmi-e++}
\!I(A\!:\!B|C)_{\omega}=\sup_{P_B}\left[\shs I(B\!:\!AC)_{Q_B\omega
Q_B}-I(B\!:\!C)_{Q_B\omega Q_B}\shs\right],\; Q_B=P_B\otimes I_{AC},\!
\end{equation}
where the suprema are over all finite rank projectors
$P_A\in\B(\H_A)$ and\break $P_B\in\B(\H_B)$ correspondingly and it is assumed that $I(X\!:\!Y)_{Q_X\omega
Q_X}=\lambda I(X\!:\!Y)_{\lambda^{-1} Q_X\omega
Q_X}$, where $\lambda=\Tr Q_X\omega_{ABC}$ \cite{CMI}.\smallskip

It is shown in \cite[Th.2]{CMI} that expressions (\ref{cmi-e+}) and
(\ref{cmi-e++}) define the same  lower semicontinuous function on the set
$\S(\H_{ABC})$ possessing all basic properties of conditional mutual
information valid in finite dimensions. In particular, the following relation (chain rule)
\begin{equation}\label{chain}
I(X\!:\!YZ|C)_{\omega}=I(X\!:\!Y|C)_{\omega}+I(X\!:\!Z|YC)_{\omega}
\end{equation}
holds for any state $\omega$ in $\S(\H_{XYZC})$ (with possible values $+\infty$ in the both sides).
To prove (\ref{chain}) is suffices to note that it holds if the systems $X,Y,Z$ and $C$ are finite-dimensional and to apply the approximating property from the second part of Theorem 2 in \cite{CMI}.

We will also use the upper bound
\begin{equation}\label{CMI-UB}
I(A\!:\!B|C)_{\omega}\leq 2\min\left\{H(\omega_A),H(\omega_B),H(\omega_{AC}),H(\omega_{BC})\right\}
\end{equation}
valid for any state $\omega_{ABC}$. It directly follows from upper bound (\ref{MI-UB}) and the expression
$I(X\!:\!Y|C)_{\omega}=I(X\!:\!YC)_{\omega}-I(X\!:\!C)_{\omega}$, $X,Y=A,B$, which is a partial case of (\ref{chain}). \smallskip

The conditional quantum mutual information is not concave or convex but the following relation
\begin{equation}\label{F-c-b}
\begin{array}{cc}
\left|\lambda
I(A\!:\!B|C)_{\rho}+(1-\lambda)I(A\!:\!B|C)_{\sigma}-I(A\!:\!B|C)_{\lambda\rho+(1-\lambda)\sigma}\right|\leq h_2(\lambda)
\end{array}
\end{equation}
holds for $\lambda\in(0,1)$ and any states $\rho_{ABC}$, $\sigma_{ABC}$ with finite $I(A\!:\!B|C)_{\rho}$, $I(A\!:\!B|C)_{\sigma}$ \cite{CHI}.
\smallskip

A finite or
countable collection $\{\rho_{i}\}$ of states
with a probability distribution $\{p_{i}\}$ is conventionally called
\textit{ensemble} and denoted $\{p_{i},\rho_{i}\}$. The state
$\bar{\rho}\doteq\sum_{i}p_{i}\rho_{i}$ is called \emph{average state} of this  ensemble. \smallskip

The \emph{Holevo quantity} of an ensemble $\{p_i,\rho_i\}_{i=1}^m$ of $\,m\leq+\infty$ quantum states is defined as
$$
\chi\left(\{p_i,\rho_i\}_{i=1}^m\right)\doteq \sum_{i=1}^m p_i H(\rho_i\|\bar{\rho})=H(\bar{\rho})-\sum_{i=1}^m p_i H(\rho_i),\quad \bar{\rho}=\sum_{i=1}^m p_i\rho_i,
$$
where the second formula is valid if $H(\bar{\rho})<+\infty$. This quantity gives the upper bound for classical information which can be obtained by applying quantum measurements to an ensemble \cite{H-73}. It plays important role in analysis of information properties of quantum systems and channels \cite{H-SCI,N&Ch,Wilde}.\smallskip

Let $\H_A=\H$ and $\,\{|i\rangle\}_{i=1}^m$ be an orthonormal basis in a $m$-dimensional Hilbert space $\H_B$. Then it is easy to show that
\begin{equation}\label{chi-rep}
\chi(\{p_i,\rho_i\}_{i=1}^m)=I(A\!:\!B)_{\hat{\omega}},\textrm{ where }\,\hat{\omega}_{AB}=\sum_{i=1}^m p_i\rho_i\otimes |i\rangle\langle i|.
\end{equation}
We will call the state $\hat{\omega}_{AB}$  in (\ref{chi-rep})  a \emph{$qc$-state} determined by the ensemble $\{p_i,\rho_i\}_{i=1}^m$.

\smallskip

By using representation (\ref{chi-rep}) it is shown in \cite{CHI} that
\begin{equation}\label{CHI-CB+}
\left|\chi(\{p_i,\rho_i\})-\chi(\{q_i,\sigma_i\})\right|\leq \varepsilon
\log \min\{\dim\H, m\}+2g(\varepsilon)
\end{equation}
for arbitrary ensembles $\,\{p_i,\rho_i\}$  and  $\,\{q_i,\sigma_i\}$ consisting of $\;m\shs$ states in $\,\S(\H)$, where $\;\varepsilon=\frac{1}{2}\sum_{i=1}^m\|\shs p_i\rho_i-q_i\sigma_i\|_1\,$ and
$\,g(\varepsilon)=(1+\varepsilon)h_2\!\left(\frac{\varepsilon}{1+\varepsilon}\right)\,$.
It is also shown that the continuity bound (\ref{CHI-CB+}) is tight and that the factor $\,2$ in  (\ref{CHI-CB+}) can be removed if  $\,\rho_i\equiv \sigma_i$.\smallskip

We will repeatedly use the following simple lemma. \smallskip
\begin{lemma}\label{sl} \emph{If $\,U$ and $\,V$ are isometries from $\H$ into $\H'$ then
$$
\|U\rho U^*-V\rho V^*\|_1\leq2\|U-V\|
$$
for any state $\rho$ in $\S(\H)$.}
\end{lemma}

\section{Special continuity bound for $I(A\!:\!B|C)$.}

Our main technical tool is the following proposition proved by simple modification of the  Alicki-Fannes-Winter method \cite{A&F,W-CB}.
\smallskip

\begin{property}\label{S-CMI-CB} \emph{Let $\rho$ and $\sigma$ be states in $\S(\H_{ABC})$\footnote{Here and in what follows systems $A,B,C,...$ are assumed infinite-dimensional.} having extensions $\hat{\rho}$ and  $\hat{\sigma}$ in $\S(\H_{ABCE})$ (i.e.  $\hat{\rho}_{ABC}=\rho$ and $\hat{\sigma}_{ABC}=\sigma$)
such that $\hat{\rho}_{AE}$ and $\hat{\sigma}_{AE}$ are finite rank states. Let $\,d\doteq\dim\left(\shs\supp\shs\hat{\rho}_{AE}\vee\supp\shs\hat{\sigma}_{AE}\right)$.\footnote{i.e. $d$ -- dimension of the minimal subspace containing the supports of $\hat{\rho}_{AE}$ and $\hat{\sigma}_{AE}$.} Then $I(A\!:\!B|C)_{\rho}$ and $I(A\!:\!B|C)_{\sigma}$ are finite and
\begin{equation}\label{S-CMI-CB+}
|I(A\!:\!B|C)_{\rho}-I(A\!:\!B|C)_{\sigma}|\leq 2\varepsilon
\log d+2g(\varepsilon),
\end{equation}
where $\;\varepsilon=\frac{1}{2}\|\shs\hat{\rho}-\hat{\sigma}\|_1\,$ and $\,g(\varepsilon )\!\doteq\!(1+\varepsilon)h_2\!\left(\frac{\varepsilon}{1+\varepsilon}\right)=(1+\varepsilon)\log(1+\varepsilon)-\varepsilon\log\varepsilon$.}

\smallskip

\emph{If $\,\hat{\rho}$ and  $\,\hat{\sigma}$ are $qc$-states with respect to the decomposition $(AE)(BC)$ then the factor $\shs2$ in the first term of (\ref{S-CMI-CB+}) can be removed. If
the function $\,\omega\mapsto I(A\!:\!B|C)_{\omega}$ is either concave or convex on the convex hull of the states $\,\rho, \sigma, \Tr_E\tau_+, \Tr_E\tau_-$, where $\tau_{\pm}=[\shs\hat{\sigma}-\hat{\rho}\shs]_\pm/\Tr[\shs\hat{\sigma}-\hat{\rho}\shs]_{\pm}$, then the factor $\shs2$ in the second term of (\ref{S-CMI-CB+}) can be removed.}
\end{property}\medskip

\emph{Proof.}  Following \cite{W-CB} introduce the state
$\,\omega^{*}=(1+\varepsilon)^{-1}(\hat{\rho}+[\shs\hat{\sigma}-\hat{\rho}\shs]_+)$ in $\S(\H_{ABCE})$. Then
\begin{equation*}
\frac{1}{1+\varepsilon}\,\hat{\rho}+\frac{\varepsilon}{1+\varepsilon}\,\tau_+=\omega^{*}=
\frac{1}{1+\varepsilon}\,\hat{\sigma}+\frac{\varepsilon}{1+\varepsilon}\,\tau_-,
\end{equation*}
where $\,\tau_+=\varepsilon^{-1}[\shs\hat{\sigma}-\hat{\rho}\shs]_+\,$
and
$\,\tau_-=\varepsilon^{-1}[\shs\hat{\sigma}-\hat{\rho}\shs]_-\,$
are states in $\S(\H_{ABCE})$. By taking partial trace we obtain
\begin{equation}\label{omega-star}
\frac{1}{1+\varepsilon}\,\rho+\frac{\varepsilon}{1+\varepsilon}\,\Tr_E\tau_+=\omega^{*}_{ABC}=
\frac{1}{1+\varepsilon}\,\sigma+\frac{\varepsilon}{1+\varepsilon}\,\Tr_E\tau_-,
\end{equation}
Since the operators
$\Tr_{BC}[\hat{\rho}-\hat{\sigma}]_{\pm}$ are supported by the $d$-dimensional subspace $\supp\shs\hat{\rho}_{AE}\vee\supp\shs\hat{\sigma}_{AE}$, basic properties of the conditional mutual information and upper bound (\ref{MI-UB}) imply
\begin{equation}\label{s-UB}
I(A\!:\!B|C)_{\omega}\leq I(A\!:\!BC)_{\omega}\leq I(AE\!:\!BC)_{\omega}\leq 2H(\omega_{AE})\leq2\log d<+\infty
\end{equation}
for $\omega=\hat{\rho},\hat{\sigma},\tau_+,\tau_-$.
By applying (\ref{F-c-b}) to the above convex decompositions of $\,\omega_{ABC}^{*}$ we obtain
$$
(1-p)\left[I(A\!:\!B|C)_{\rho}-I(A\!:\!B|C)_{\sigma}\right]\leq p
\left[I(A\!:\!B|C)_{\tau_-}
-I(A\!:\!B|C)_{\tau_+}\right]+2\shs
h_2(p)
$$
and
$$
(1-p)\left[I(A\!:\!B|C)_{\sigma}-I(A\!:\!B|C)_{\rho}\right]\leq p
\left[I(A\!:\!B|C)_{\tau_+}-
I(A\!:\!B|C)_{\tau_-}\right]+2\shs h_2(p).
$$
where $p=\frac{\varepsilon}{1+\varepsilon}$. These inequalities, upper bound (\ref{s-UB}) and  nonnegativity of $I(A\!:\!B|C)$  imply (\ref{S-CMI-CB+}).

If $\,\hat{\rho}$ and  $\,\hat{\sigma}$ are $qc$-states with respect to the decomposition $(AE)(BC)$ then the
above states $\tau_+$ and $\tau_-$ are $qc$-states as well. So, by using (\ref{MI-UB+}) instead of (\ref{MI-UB}) we can obtain $\log d$ instead
$2\log d$ in the right side of (\ref{s-UB}).\smallskip

The assertion concerning possibility to remove the factor $\shs2$ in the second term of (\ref{S-CMI-CB+}) is easily derived from the above arguments.
$\square$

\section{Continuity bounds for the output Holevo\\ quantity depending on input dimension}

In analysis of information properties of quantum channels we have to consider the output Holevo quantity of a given channel $\Phi:A\rightarrow B$ corresponding to an ensemble $\{p_i,\rho_i\}$ of input quantum states, i.e. the quantity
$$
\chi_{\Phi}(\{p_i,\rho_i\})\doteq \sum_{i} p_i H(\Phi(\rho_i)\|\Phi(\bar{\rho}))=H(\Phi(\bar{\rho}))-\sum_{i}p_i H(\Phi(\rho_i)),
$$
where $\bar{\rho}=\sum_{i} p_i\rho_i$ and the second formula is valid if $H(\Phi(\bar{\rho}))<+\infty$. If the state $\,\bar{\rho}\,$ is supported by some finite-dimensional subspace $\H^0_A\subseteq\H_A$ then the value $\chi_{\Phi}(\{p_i,\rho_i\})$ is finite and does not exceed $\log\dim\H^0_A$ (despite possible infinite values of the output entropies $H(\Phi(\rho_i)$).

We will consider the quantity $\chi_{\Phi}(\{p_i,\rho_i\})$ as a function of a pair (channel $\Phi$, input ensemble $\{p_i,\rho_i\}$) assuming that the set of all channels is equipped with the Bures distance (see Section 1). If $\Phi$ and $\Psi$ are quantum channels from arbitrary system $A$ to finite-dimensional system $B$ then it follows from the continuity bound (\ref{CHI-CB+}) that
$$
\left|\chi_{\Phi}(\{p_i,\rho_i\})-\chi_{\Psi}(\{q_i,\sigma_i\})\right|\leq \varepsilon
\log \min\{d_B, m\}+2g(\varepsilon)
$$
for any ensembles $\,\{p_i,\rho_i\}$  and  $\,\{q_i,\sigma_i\}$ consisting of $\;m\shs$ states in $\,\S(\H_A)$, where $\;\varepsilon=\frac{1}{2}\sum_{i=1}^m\|\shs p_i\rho_i-q_i\sigma_i\|_1+\|\Phi-\Psi\|\,$ and
$\,g(\varepsilon)=(1+\varepsilon)h_2\!\left(\frac{\varepsilon}{1+\varepsilon}\right)$.\footnote{$\|\Upsilon\|$ denotes the operator norm of the map $\Upsilon:\T(\H_A)\rightarrow\T(\H_B)$.}\smallskip

The following proposition gives continuity bounds for $\chi_{\Phi}\left(\{p_i,\rho_i\}\right)$  depending  on the  dimension of input subspace containing states of ensembles.\smallskip

\begin{property}\label{HQ-CB} \emph{Let $\,\Phi:A\rightarrow B$ and $\,\Psi:A\rightarrow B$ be arbitrary quantum channels.
Let $\,\{p_i,\rho_i\}$  and  $\,\{q_i,\sigma_i\}$ be ensembles of states in $\,\S(\H_A)$ supported by some finite-dimensional subspace $\H^0_A\subseteq\H_A$. Then
\begin{equation}\label{HQ-CB+}
\left|\chi_{\Phi}(\{p_i,\rho_i\})-\chi_{\Psi}(\{q_i,\sigma_i\})\right|\leq \varepsilon
\log d_A+\varepsilon
\log 2+2g(\varepsilon),
\end{equation}
where $\,d_A\doteq\dim\H^0_A$, $\,\varepsilon=\frac{1}{2}\sum_{i}\|\shs p_i\rho_i-q_i\sigma_i\|_1+\beta(\Phi,\Psi),\,g(\varepsilon)=(1+\varepsilon)h_2\!\left(\frac{\varepsilon}{1+\varepsilon}\right)$.}

\smallskip

\emph{If $\,\Phi=\Psi$ then the summand $\shs\varepsilon
\log 2$ in (\ref{HQ-CB+}) can be removed.
If  $\,\Phi(\rho_i)\equiv \Psi(\sigma_i)$  then the factor $\,2$ in (\ref{HQ-CB+}) can be removed.}
\smallskip

\emph{Continuity bound (\ref{HQ-CB+}) is tight in the cases $\,\Phi=\Psi$ and $\,\{p_i,\rho_i\}=\{q_i,\sigma_i\}$. The Bures distance $\beta(\Phi,\Psi)$ in (\ref{HQ-CB+}) can be replaced by $\|\Phi-\Psi\|^{1/2}_{\diamond}$.}
\end{property}\medskip

\emph{Proof.} Let $E$ be a common environment for the channels $\Phi$ and $\Psi$, so that representation (\ref{c-S-r}) holds with some isometries
$V_{\Phi}$ and $V_{\Psi}$ from $\H_A$ into $\H_{BE}$.
Consider the $qc$-states
$$
\hat{\rho}=\sum_{i} p_iV_{\Phi}\rho_iV_{\Phi}^*\otimes |i\rangle\langle i|\quad
\textrm{and}\quad \hat{\sigma}=\sum_{i}q_iV_{\Psi}\sigma_iV_{\Psi}^*\otimes |i\rangle\langle i|
$$
in $\S(\H_{BEC})$, where $\{|i\rangle\}$ is a basic in $\H_C$. It follows from (\ref{chi-rep}) that
$$
\chi_{\Phi}(\{p_i,\rho_i\})=I(B\!:\!C)_{\hat{\rho}}\quad \textrm{and} \quad \chi_{\Psi}(\{q_i,\sigma_i\})=I(B\!:\!C)_{\hat{\sigma}}.
$$
Lemma \ref{sl} implies
\begin{equation}\label{norm+}
 \|\shs\hat{\rho}-\hat{\sigma}\|_1\leq \sum_{i}\|p_i\rho_i-q_i\sigma_i\|_1+2\|V_{\Phi}-V_{\Psi}\|.
\end{equation}
Since the states  $\hat{\rho}_{BE}$ and $\hat{\sigma}_{BE}$
are supported by the subspace $V_{\Phi}\H^0_A\vee V_{\Psi}\H^0_A$ of $\H_{BE}$ having dimension $\leq2d_A$,
Proposition \ref{S-CMI-CB} and (\ref{norm+}) imply (\ref{HQ-CB+}). If $\Phi=\Psi$ then the above states
$\hat{\rho}_{BE}$ and $\hat{\sigma}_{BE}$ are supported by the $d_A$-dimensional subspace  $V_{\Phi}\H^0_A=V_{\Psi}\H^0_A$.

The tightness of continuity bound (\ref{HQ-CB+}) in the case $\,\Phi=\Psi$ follows from the tightness of the continuity bound (\ref{CHI-CB+}), see Remark 3 in \cite{CHI}.\smallskip

The tightness of continuity bound (\ref{HQ-CB+}) in the case $\,\{p_i,\rho_i\}=\{q_i,\sigma_i\}$  follows from the tightness of continuity bound (\ref{HC-CB}) for the Holevo capacity in Section 3 (which is derived from (\ref{HQ-CB+})). It can be directly shown by using the erasure channels $\Phi_{1/2}$ and $\Phi_{1/2-x}$ (see the proof of Proposition \ref{cap-tcb}).\smallskip

If $\,\Phi(\rho_i)\equiv \Psi(\sigma_i)$ then all the states $\hat{\rho}_{BC},\hat{\sigma}_{BC},[\tau_{-}]_{BC},[\tau_{+}]_{BC}$  are $qc$-states determined by ensembles with the same set of states and hence the function $\omega\mapsto I(B\!:\!C)_{\omega}$ is concave on the convex hull of these states \cite[Th.13.3.3]{Wilde}. So, the assertion concerning possibility to remove the factor $2$ in the second term of (\ref{HQ-CB+})  follows from the corresponding  assertion of Proposition \ref{S-CMI-CB}.\smallskip

The last assertion of the proposition follows from the right inequality in (\ref{DB-rel}) and monotonicity of the function $g(x)$. $\square$

\section{Continuity bounds for output conditional mutual information depending on input dimension}

Quantum mutual information and its conditional version play basic role in analysis of informational properties of quantum channels (see Section 5).
In this section we will explore continuity properties of the conditional mutual information at output of a channel acting on one subsystem of a tripartite system, i.e. the quantity $I(B\!:\!D|C)_{\Phi\otimes\id_{CD}(\rho)}$, where $\Phi:A\rightarrow B$ is an arbitrary channel, $C,D$ are any systems and $\rho$ is a state in $\S(\H_{ADC})$. If the marginal state $\,\rho_A\doteq\Tr_{CD}\rho\,$ has finite rank then upper bound (\ref{CMI-UB}) and monotonicity of the conditional mutual information under local channels show that this quantity does not exceed $2\log \mathrm{rank}\rho_A$ (despite possible infinite values of all marginal entropies of the state $\Phi\otimes\id_{CD}(\rho)$).

We will obtain tight continuity bound for the function $$(\Phi, \rho)\mapsto I(B\!:\!D|C)_{\Phi\otimes\id_{CD}(\rho)}$$
assuming that the set of all channels from $A$ to $B$ is equipped with the Bures distance (equivalent to the metric induced by the norm of complete boundedness, see Section 1). We will also obtain tight continuity bound for the function
$$
\Phi\mapsto I(B^n\!:\!D|C)_{\Phi^{\otimes n}\otimes\id_{CD}(\rho)},\quad \textrm{where }\, \rho\in\S(\H_{A^nCD}),
$$
for any natural $\shs n\,$ depending on the ranks of the states $\rho_{A_1},...,\rho_{A_n}$.

\subsection{The function $(\Phi, \rho)\mapsto I(B\!:\!D|C)_{\Phi\otimes\id_{CD}(\rho)}$}

\begin{property}\label{MI-CB} \emph{Let $\,\Phi:A\rightarrow B$ and $\,\Psi:A\rightarrow B$ be arbitrary quantum channels and $\,C,D$ be any systems.
Let $\rho$ and $\sigma$ be states in $\S(\H_{ACD})$ such that the states $\rho_A$ and $\sigma_A$ are supported by some finite-dimensional subspace $\H^0_A\subseteq\H_A$. Then
\begin{equation}\label{MI-CB+}
\!|I(B\!:\!D|C)_{\Phi\otimes\id_{CD}(\rho)}-I(B\!:\!D|C)_{\Psi\otimes\id_{CD}(\sigma)}|\leq 2\varepsilon\log d_A+2\varepsilon
\log 2+2g(\varepsilon),\!
\end{equation}
where $\,d_A\doteq\dim\H^0_A$, $\,\varepsilon=\frac{1}{2}\|\shs\rho-\sigma\|_1+\beta(\Phi,\Psi)$ and $\,g(\varepsilon)=(1+\varepsilon)h_2\!\left(\frac{\varepsilon}{1+\varepsilon}\right)$.}
\smallskip

\emph{If $\,\Phi=\Psi$ then the summand $\shs2\varepsilon
\log 2$ in (\ref{MI-CB+}) can be removed.}
\smallskip

\emph{Continuity bound (\ref{MI-CB+}) is tight in the both cases $\,\Phi=\Psi$ and $\,\rho=\sigma$ even for trivial $C$ when $I(B\!:\!D|C)=I(B\!:\!D)$. The Bures distance $\beta(\Phi,\Psi)$ in (\ref{MI-CB+}) can be replaced by $\|\Phi-\Psi\|^{1/2}_{\diamond}$.}
\end{property}\medskip

\emph{Proof.} Let $E$ be a common environment for the channels $\Phi$ and $\Psi$, so that representation (\ref{c-S-r}) holds with some isometries
$V_{\Phi}$ and $V_{\Psi}$ from $\H_A$ into $\H_{BE}$. Then
$$
\hat{\rho}=\!V_{\Phi}\otimes I_{CD}\shs\rho\shs V_{\Phi}^*\!\otimes I_{CD}\;\;
\textrm{and}\;\; \hat{\sigma}=\!V_{\Psi}\otimes I_{CD}\shs\sigma\shs V_{\Psi}^*\!\otimes I_{CD}
$$
are extensions of the states $\Phi\otimes\id_{CD}(\rho)$
and $\Psi\otimes\id_{CD}(\sigma)$. Lemma \ref{sl} implies
\begin{equation}\label{norm}
 \|\shs\hat{\rho}-\hat{\sigma}\|_1\leq \|\shs\rho-\sigma\|_1+2\|V_{\Phi}-V_{\Psi}\|.
\end{equation}
Since the states  $\hat{\rho}_{BE}=V_{\Phi}\rho_AV_{\Phi}^*$ and $\hat{\sigma}_{BE}=V_{\Psi}\sigma_AV_{\Psi}^*$
are supported by the subspace $V_{\Phi}\H^0_A\vee V_{\Psi}\H^0_A$ of $\H_{BE}$ having dimension $\leq 2d_A$,
Proposition \ref{S-CMI-CB} and inequality (\ref{norm}) imply (\ref{MI-CB+}). If $\Phi=\Psi$ then the above states
$\hat{\rho}_{BE}$ and $\hat{\sigma}_{BE}$ are supported by the $d_A$-dimensional subspace  $V_{\Phi}\H^0_A=V_{\Psi}\H^0_A$.

The tightness of continuity bound (\ref{MI-CB+}) in the case $\Phi=\Psi$ follows from Corollary 1 in \cite{CHI}.\smallskip

The tightness of continuity bound (\ref{MI-CB+}) in the case $\rho=\sigma$ follows from the tightness of continuity bound (\ref{QC-CB}) for the quantum capacity in Section 4 (derived from (\ref{MI-CB+})). It can be directly shown by using the erasure channels $\Phi_{1/2}$ and $\Phi_{1/2-x}$ (see the proof of Proposition \ref{cap-tcb}).\smallskip

The last assertion of the proposition follows from the right inequality in (\ref{DB-rel}) and monotonicity of the function $g(x)$.$\square$

\subsection{The function $\Phi\mapsto I(B^n\!:\!D|C)_{\Phi^{\otimes n}\otimes\id_{CD}(\rho)}$}

The following proposition is a $d_A$-version of Lemma 2 in \cite{CHI} proved by using the Leung-Smith telescopic trick from \cite{L&S}.\smallskip

\begin{property}\label{omi} \emph{Let $\,\Phi:A\rightarrow B$ and $\,\Psi:A\rightarrow B$ be arbitrary quantum channels, $C,D$ be any systems and $\,n\in\mathbb{N}$. Let $\,\rho\shs$ be a state in $\,\S(\H^{\otimes n}_{A}\otimes\H_{CD})$ such that $\,\rho_{A_1},...,\rho_{A_n}$ are a finite rank states. Then
\begin{equation*}
\left|I(B^n\!:\!D|C)_{\Phi^{\otimes n}\otimes\id_{CD}(\rho)}-I(B^n\!:\!D|C)_{\Psi^{\otimes n}\otimes\id_{CD}(\rho)}\right|\leq 2n(\varepsilon
\log (2d_A) + g(\varepsilon)),
\end{equation*}
where $\,\varepsilon=\beta(\Phi,\Psi)$ and $\,d_A\doteq \left[\prod_{k=1}^n\mathrm{rank}\rho_{A_k}\right]^{1/n}$.}\smallskip

\emph{This continuity bound is tight even for trivial $C$ (for each given $n$ and large $d_A$). The Bures distance $\beta(\Phi,\Psi)$ in it  can be replaced by $\|\Phi-\Psi\|^{1/2}_{\diamond}$.}
\end{property}\medskip

\begin{remark}\label{omi-r}
If $\,\dim\H_A<+\infty\,$ then one can take $\,d_A\doteq\dim\H_A$.
\end{remark}\medskip

\emph{Proof.} Let $E$ be a common environment for the channels $\Phi$ and $\Psi$, so that Stinespring representations (\ref{c-S-r}) hold with some isometries
$V_{\Phi}$ and $V_{\Psi}$ from $\H_A$ into $\H_{BE}$. By Theorem 1 in \cite{Kr&W} we may assume that $\|V_{\Phi}-V_{\Psi}\|=\beta(\Phi,\Psi)$.

Consider the states
$$
\sigma_k=\Phi^{\otimes k}\otimes\Psi^{\otimes (n-k)}\otimes\id_{CD}(\rho),\quad k=0,1,...,n.
$$
We have
\begin{equation}\label{tel}
\!\!\!\!\!\begin{array}{c}
\displaystyle \left|I(B^n\!:\!D|C)_{\sigma_n}\!-I(B^n\!:\!D|C)_{\sigma_0}\right|\displaystyle=
\left|\sum_{k=1}^n I(B^n\!:\!D|C)_{\sigma_k}\!-I(B^n\!:\!D|C)_{\sigma_{k-1}}\right|\\ \leq \displaystyle \sum_{k=1}^n \left|I(B^n\!:\!D|C)_{\sigma_k}\!-I(B^n\!:\!D|C)_{\sigma_{k-1}}\right|.
\end{array}\!\!\!
\end{equation}
By using the chain rule (\ref{chain}) we obtain for each $k$ (cf.\cite{L&S})
\begin{equation}\label{tel+}
\!\!\begin{array}{ll}
I(B^n\!:\!D|C)_{\sigma_k}\!-I(B^n\!:\!D|C)_{\sigma_{k-1}}&\!\!\!\!\!= I(B_1...B_{k-1}B_{k+1}...B_n \!:\!D|C)_{\sigma_k}\\\\& \!\!\!\!\!+\,I(B_k\!:\!D|B_1...B_{k-1}B_{k+1}...B_nC)_{\sigma_k}\\\\&\!\!\!\!\!-\,
I(B_1...B_{k-1}B_{k+1}...B_n \!:\!D|C)_{\sigma_{k-1}}\\\\&\!\!\!\!\!-\,I(B_k\!:\!D|B_1...B_{k-1}B_{k+1}...B_nC)_{\sigma_{k-1}}\\\\&\!\!\!\!\!=
I(B_k\!:\!D|B_1...B_{k-1}B_{k+1}...B_nC)_{\sigma_k}\\\\&\!\!\!\!\!-\,
I(B_k\!:\!D|B_1...B_{k-1}B_{k+1}...B_nC)_{\sigma_{k-1}},\!\!\!
\end{array}
\end{equation}
where it was  used that $\Tr_{B_k}\sigma_k=\Tr_{B_k}\sigma_{k-1}$. Note that the finite rank of the states $\,\rho_{A_1},...,\rho_{A_n}$  and monotonicity of the conditional mutual information under local channels guarantee finiteness of all the terms in (\ref{tel}) and (\ref{tel+}).

To estimate the last difference in (\ref{tel+}) consider  the states
$$
\hat{\sigma}_k=\Tr_{E^n\setminus E_k} W_k\otimes V^k_{\Phi}\otimes I_{CD} \,\rho\; W^*_k\otimes [V^k_{\Phi}]^*\otimes I_{CD}
$$
and
$$
\hat{\sigma}_{k-1}=\Tr_{E^n\setminus E_k} W_k\otimes V^k_{\Psi}\otimes I_{CD} \,\rho\; W^*_k\otimes [V^k_{\Psi}]^*\otimes I_{CD}
$$
in $\S(\H_{B^nCDE_k})$, where $\,W_k=V^1_{\Phi}\otimes\ldots \otimes V^{k-1}_{\Phi}\otimes V^{k+1}_{\Psi}\otimes\ldots \otimes V^{n}_{\Psi}\,$ is an isometry from $\H_{A^n\setminus A_k}$ into $\H_{[BE]^n\setminus [BE]_k}$, $V^k_{\Phi}\cong V_{\Phi}$ and $V^k_{\Psi}\cong V_{\Psi}$ are isometries from $\H_{A_k}$ into $\H_{B_kE_k}$. It follows from (\ref{c-S-r}) that these states
are extensions of the states $\sigma_k$ and $\sigma_{k-1}$, i.e. $\Tr_{E_k}\hat{\sigma}_k=\sigma_k$ and $\Tr_{E_k}\hat{\sigma}_{k-1}=\sigma_{k-1}$. Since
$$
[\hat{\sigma}_k]_{B_kE_k}=V^k_{\Phi}\rho_{A_k} [V^k_{\Phi}]^*\quad \textrm{and} \quad [\hat{\sigma}_{k-1}]_{B_kE_k}=V^k_{\Psi}\rho_{A_k} [V^k_{\Psi}]^*,
$$
Proposition \ref{S-CMI-CB} implies
\begin{equation}\label{tel++}
|I(B_k\!:\!D|X)_{\sigma_k}-
I(B_k\!:\!D|X)_{\sigma_{k-1}}|\leq 2\varepsilon' \log (2\shs\mathrm{rank}\rho_{A_k})+2g(\varepsilon'),
\end{equation}
where $X=B_1...B_{k-1}B_{k+1}...B_nC$ and $\,\varepsilon'=\frac{1}{2}\|\hat{\sigma}_{k}-\hat{\sigma}_{k-1}\|_{1}$. Since the trace norm does not increase under partial trace, by using Lemma \ref{sl} we obtain
$$
\!\!\!\begin{array}{rl}
\varepsilon'\!\!\!\!& \leq \textstyle\frac{1}{2}\|W_k\otimes V^k_{\Phi}\otimes I_{CD} \,\rho\; W^*_k\otimes [V^k_{\Phi}]^*\!\otimes I_{CD}-W_k\otimes V^k_{\Psi}\otimes I_{CD} \,\rho\; W^*_k\otimes [V^k_{\Psi}]^*\!\otimes I_{CD}\|_{1}\\\\& \leq\|W_k\otimes V^k_{\Phi}\otimes I_{CD}-W_k\otimes V^k_{\Psi}\otimes I_{CD}\|=\|V_{\Phi}-V_{\Psi}\|=\beta(\Phi,\Psi)=\varepsilon.
\end{array}
$$
Hence, it follows from  (\ref{tel+}) and (\ref{tel++}) that
$$
\left|I(B^n\!:\!D|C)_{\sigma_k}-I(B^n\!:\!D|C)_{\sigma_{k-1}}\right|\leq  2\varepsilon \log (2\shs\mathrm{rank}\rho_{A_k})+2g(\varepsilon).
$$
This and (\ref{tel}) imply the required inequality  (since $\Phi^{\otimes n}\otimes\id_{CD}(\rho)=\sigma_n$ and $\Psi^{\otimes n}\otimes\id_{CD}(\rho)=\sigma_0$).\smallskip

The tightness of the continuity bound in Proposition \ref{omi} follows from the tightness of continuity bound (\ref{MI-CB+}), since
for arbitrary channel $\Phi:A\rightarrow B$, any system $D$ and a state $\rho\in\S(\H_{AD})$ we have
$$
I(B^n\!:\!D^n)_{\Phi^{\otimes n}\otimes\id_{D^n}(\rho^{\otimes n})}=nI(B\!:\!D)_{\Phi\otimes\id_{D}(\rho)}.
$$
The last assertion of the proposition follows from the right inequality in (\ref{DB-rel}) and monotonicity of the function $g(x)$.
$\square$

\section{Continuity bounds for basic capacities of channels with finite input dimension.}

Continuity bounds for basic capacities of quantum channels with finite output dimension $d_B$ are obtained by Leung and Smith in \cite{L&S}. The main term in all these bounds has the form $C\varepsilon\log d_B$ for some constant $C$, where $\varepsilon$ is a distance between two channels (the  diamond norm of their difference). These continuity bounds are essentially refined in \cite{CHI} by using modification of the Leung-Smith approach (consisting in using the conditional mutual information instead of the conditional entropy).\smallskip

In this section we consider quantum channels with finite input  dimension\footnote{Channels with finite input dimension and infinite output dimension may appear as subchannels of "real" infinite-dimensional channels if we use for coding information only states supported by some finite-dimensional subspace of the input space.} $d_A$ and  obtain
continuity bounds for basic capacities of such channels with the main term $C\varepsilon\log d_A$, where $\varepsilon$ is the Bures distance between quantum channels described in Section 1. \smallskip

The \emph{Holevo capacity} of a quantum channel
$\Phi:A\rightarrow B$  is
defined as follows
\begin{equation}\label{HC-def}
\bar{C}(\Phi)=\sup_{\{p_i,\rho_i\}\in \mathcal{E}(\H_A)}\chi_{\Phi}(\{p_i,\rho_i\}),
\end{equation}
where $\mathcal{E}(\H_A)$ is the set of all ensembles of input states. This quantity is closely related to the classical capacity of a quantum channel (see below).\smallskip

By the Holevo-Schumacher-Westmoreland theorem  the \emph{classical capacity} of
a  channel $\Phi:A\rightarrow B$  is given by
the  regularized expression
\begin{equation}\label{CC-def}
C(\Phi)=\lim_{n\rightarrow +\infty }n^{-1}\bar{C}(\Phi^{\otimes n}).
\end{equation}

The \emph{classical
entanglement-assisted capacity} of a quantum channel determines an ultimate rate of transmission of classical information when an entangled state between the input and the output of a channel is used as an
additional resource (see details in \cite{H-SCI,Wilde}). By the Bennett-Shor-Smolin-Thaplyal
theorem the classical
entanglement-assisted capacity of a  channel
$\Phi:A\rightarrow B$ is given by the expression
\begin{equation}\label{EAC-def}
C_{\mathrm{ea}}(\Phi)=\sup_{\rho \in
\mathfrak{S}(\mathcal{H}_A)}I(\Phi, \rho),
\end{equation}
in which $\shs I(\Phi, \rho)\shs$ is the quantum mutual information of a channel $\Phi$ at a state $\rho$ defined as follows
\begin{equation*}
 I(\Phi,\rho)=I(B\!:\!R)_{\Phi\otimes\mathrm{Id}_{R}(\hat{\rho})},
\end{equation*}
where $\mathcal{H}_R\cong\mathcal{H}_A$ and $\hat{\rho}$
is a pure state in $\S(\H_{AR})$ such that $\hat{\rho}_{A}=\rho$  \cite{H-SCI,Wilde}.

\smallskip

The \emph{quantum capacity} of a channel characterizes ultimate rate of transmission of quantum information (quantum states) through a channel.
By the Lloyd-Devetak-Shor  theorem  the quantum capacity of
a  channel $\Phi:A\rightarrow B$  is given by
the  regularized expression
\begin{equation}\label{QC-def}
Q(\Phi)=\lim_{n\rightarrow +\infty }n^{-1}\bar{Q}(\Phi^{\otimes n}),
\end{equation}
where $\bar{Q}(\Phi)$ is the maximal value of the coherent information $I_c(\Phi,\rho)\doteq H(\Phi(\rho))-H(\widehat{\Phi}(\rho))$ over all input states $\rho\in \S(\H_A)$ (here $\widehat{\Phi}$ is a complementary channel to the channel $\Phi$ defined in (\ref{c-channel})) \cite{H-SCI,Wilde}. \smallskip

The \emph{private capacity} is the capacity of a channel for classical communication with the
additional requirement that almost no information is sent to the environment. By the Devetak theorem
the private capacity of a  channel $\Phi:A\rightarrow B$  is given by
the regularized expression
\begin{equation}\label{PCC-def}
C_\mathrm{p}(\Phi)=\lim_{n\rightarrow +\infty }n^{-1}\bar{C}_\mathrm{p}(\Phi^{\otimes n}),
\end{equation}
where
\begin{equation}\label{PHC-def}
\bar{C}_\mathrm{p}(\Phi)=\sup_{\{p_i,\rho_i\}\in \mathcal{E}(\H_A)}\left[\chi_{\Phi}(\{p_i,\rho_i\})-\chi_{\widehat{\Phi}}(\{p_i,\rho_i\})\right]
\end{equation}
is the private analog of the Holevo capacity (here $\widehat{\Phi}$ is a complementary channel to the channel $\Phi$ defined in (\ref{c-channel})) \cite{H-SCI,Wilde}.
\medskip

Now we consider continuity bounds for all the above capacities depending on the input dimension. For the entanglement-assisted classical capacity
the tight continuity bound
\begin{equation*}
\left|\shs C_{\mathrm{ea}}(\Phi)-C_{\mathrm{ea}}(\Psi)\right|\leq 2\varepsilon
\log d_A +2g(\varepsilon),
\end{equation*}
where $\,\varepsilon=\frac{1}{2}\|\Phi-\Psi\|_{\diamond}$, is obtained in \cite[Pr.6]{CHI}. For others capacities tight and close-to-tight
continuity bounds depending on input dimension are presented in the following proposition, in which the Bures distance $\beta(\Phi,\Psi)$
described in Section 1 is used as a measure of the difference $\Phi-\Psi$ instead of $\frac{1}{2}\|\Phi-\Psi\|_{\diamond}$.

\smallskip

\begin{property}\label{cap-tcb} \emph{Let $\,\Phi$ and $\,\Psi$ be quantum  channels from finite-dimensional system $A$ to arbitrary system $B$.\footnote{We assume that expressions (\ref{HC-def})-(\ref{PHC-def}) remain valid in the case $\dim\H_B=+\infty$.}  Then
\begin{equation}\label{HC-CB}
|\shs \bar{C}(\Phi)-\bar{C}(\Psi)|\leq \varepsilon
\log d_A + \varepsilon
\log 2 + 2g(\varepsilon),\quad\;\;
\end{equation}
\begin{equation}\label{CC-CB}
\left|\shs C(\Phi)-C(\Psi)\right|\leq 2\varepsilon
\log d_A +2\varepsilon
\log 2+2g(\varepsilon),\;\;
\end{equation}
\begin{equation}\label{QC-CB}
\left|\shs Q(\Phi)-Q(\Psi)\right|\leq 2\varepsilon
\log d_A +2\varepsilon
\log 2+2g(\varepsilon),\;\;
\end{equation}
\begin{equation}\label{PHC-CB}
|\shs\bar{C}_{\mathrm{p}}(\Phi)-\bar{C}_{\mathrm{p}}(\Psi)|\leq 2\varepsilon
\log d_A+2\varepsilon
\log 2 +2g(\varepsilon),
\end{equation}
\begin{equation}\label{PCC-CB}
\left|\shs C_{\mathrm{p}}(\Phi)-C_{\mathrm{p}}(\Psi)\right|\leq 4\varepsilon
\log d_A +4\varepsilon
\log 2+4g(\varepsilon),
\end{equation}
where  $d_A\doteq\dim\H_A$, $\,\varepsilon=\beta(\Phi,\Psi)\,$  and $\;g(\varepsilon)=(1+\varepsilon)h_2\!\left(\frac{\varepsilon}{1+\varepsilon}\right)$.} \medskip

\emph{The continuity bounds (\ref{HC-CB}),(\ref{QC-CB}) and (\ref{PHC-CB}) are tight. In all the inequalities (\ref{HC-CB})-(\ref{PCC-CB}) the Bures  distance $\beta(\Phi,\Psi)$ can be replaced by  $\,\|\Phi-\Psi\|^{1/2}_{\diamond}$.}
\end{property}\medskip

\emph{Proof.} Continuity bound (\ref{HC-CB}) directly follows from Proposition \ref{HQ-CB} and the definition of the Holevo capacity.

Continuity bound (\ref{CC-CB}) is  obtained by using  Lemma 12 in \cite{L&S}, representation (\ref{chi-rep}) and Proposition \ref{omi}.

To prove  continuity bound (\ref{QC-CB}) note that the coherent information can be represented as follows
$$
I_c(\Phi,\rho)= I(B\!:\!R)_{\Phi\otimes\id_R}(\hat{\rho})-H(\rho),
$$
where $\hat{\rho}$ is a purification in $\S(\H_{AR})$ of a state $\rho$. Hence for arbitrary quantum channels $\Phi$ and $\Psi$, any $n$ and a state $\rho$ in $\S(\H^{\otimes n}_{A})$ we have
$$
I_c(\Phi^{\otimes n}\!,\rho)-I_c(\Psi^{\otimes n}\!,\rho)=I(B^n\!:\!R^n)_{\Phi^{\otimes n}\otimes\id_{R^n}}(\hat{\rho})-I(B^n\!:\!R^n)_{\Psi^{\otimes n}\otimes\id_{R^n}}(\hat{\rho})
$$
where $\hat{\rho}$ is a purification  in $\S(\H^{\otimes n}_{AR})$ of the state $\rho$. This representation, Proposition \ref{omi} and Lemma 12 in \cite{L&S} imply (\ref{QC-CB}).

Continuity bound (\ref{PHC-CB}) is  obtained by using  Proposition \ref{HQ-CB} twice and by noting that $\beta(\widehat{\Phi},\widehat{\Psi})=\beta(\Phi,\Psi)$.

To prove  continuity bound (\ref{PCC-CB}) note that representation (\ref{chi-rep}) implies
\begin{equation}\label{PHC-def}
\bar{C}_\mathrm{p}(\Phi^{\otimes n})=\sup_{\hat{\rho}} \left[I(B^n\!:\!C)_{\Phi^{\otimes n}\otimes\id_{C}(\hat{\rho})}-I(E^n\!:\!C)_{\widehat{\Phi}^{\otimes n}\otimes\id_{C}(\hat{\rho})}\right],
\end{equation}
where the supremum is over all $qc$-states in $A^nC$. Since $\beta(\widehat{\Phi},\widehat{\Psi})=\beta(\Phi,\Psi)$,  inequality (\ref{PCC-CB}) is obtained by using Proposition \ref{omi} twice and Lemma 12 in \cite{L&S}.\smallskip

To show the tightness of continuity bounds (\ref{HC-CB}),(\ref{QC-CB}) and (\ref{PHC-CB}) consider the family of erasure channels
$$
\Phi_p(\rho)=\left[\begin{array}{cc}
(1-p)\rho &  0 \\
0 &  p\Tr\rho
\end{array}\right], \quad p\in[0,1].
$$
from $d$-dimensional system $A$ to $(d+1)$-dimensional system $B$. It is well known (see \cite{H-SCI,Wilde}) that
\begin{equation}\label{ecc}
  C(\Phi_p)=\bar{C}(\Phi_p)=(1-p)\log d
\end{equation}
and
\begin{equation}\label{ecc+}
  Q(\Phi_p)=C_\mathrm{p}(\Phi_p)=\bar{C}_\mathrm{p}(\Phi_p)=\max\{(1-2p)\log d, 0\}.
\end{equation}
By writing the  channel $\Phi_p$ as the map $\rho\mapsto (1-p)\rho\oplus [p\Tr\rho]|\psi\rangle\langle\psi|$ from $\T(\H_A)$ to $\T(\H_A\oplus\H_{\psi})$, where $\H_{\psi}$ is the space generated by $|\psi\rangle$, we see that the isometry
$$
V_p:|\varphi\rangle\mapsto\sqrt{1-p}|\varphi\rangle\otimes|\psi\rangle\oplus\sqrt{p}|\psi\rangle\otimes|\varphi\rangle
$$
from $\H_A$ into $\H_{BE}$, where $\H_{E}=\H_{B}=\H_A\oplus\H_{\psi}$, is a Stinespring isometry for $\Phi_p$, i.e.
$\Phi_p(\rho)=\Tr_E V_{p}\rho V^*_{p}$, for each $p$. Direct calculation shows that
\begin{equation}\label{dV}
 \|V_{1/2-x}-V_{1/2}\|=\sqrt{2-\sqrt{1-2x}-\sqrt{1+2x}}=x+o(x)\quad (x\rightarrow0).
\end{equation}
It follows from (\ref{ecc}) and (\ref{ecc+}) that $\;\bar{C}(\Phi_{1/2-x})-\bar{C}(\Phi_{1/2})=x\log d\;$ and that
$$
Q(\Phi_{1/2-x})-Q(\Phi_{1/2})=\bar{C}_\mathrm{p}(\Phi_{1/2-x})-\bar{C}_\mathrm{p}(\Phi_{1/2})=2x\log d
$$
Since (\ref{dV}) implies $\beta(\Phi_{1/2-x},\Phi_{1/2})\leq x+o(x)$ for small $x$, we see that continuity bounds (\ref{HC-CB}),(\ref{QC-CB}) and (\ref{PHC-CB}) are tight (for large $d_A$). \smallskip

The last assertion of the proposition follows from the right inequality in (\ref{DB-rel}) and monotonicity of the function $g(x)$. $\square$

\medskip

I am grateful to A.S.Holevo and G.G.Amosov for useful discussion. 

The research is funded by the grant of Russian Science Foundation
(project No 14-21-00162).

\end{document}